\documentclass[final,1p,times]{elsarticle}

\usepackage{amssymb}
\usepackage{amsthm}
\usepackage{graphicx}
\usepackage{dcolumn}
\usepackage{bm}
\usepackage{siunitx}
\usepackage{textcomp}
\usepackage[utf8]{inputenc}
\usepackage[T1]{fontenc}
\usepackage{mathptmx}
\usepackage{xcolor}
\usepackage{gensymb}
\usepackage[english]{babel}
\usepackage{tabularx}
\definecolor{modified}{RGB}{0,176,80}
\usepackage{lipsum}
\usepackage{amsmath}
\usepackage[colorlinks=true, allcolors=blue]{hyperref}

\usepackage{tikz}
\usepackage{booktabs,adjustbox}
\usepackage[flushleft]{threeparttable}

\newcounter{magicrownumbers}

\DeclareUnicodeCharacter{2212}{-}
\journal{Spectrochimica Acta Part B: Atomic Spectroscopy}

\begin{document}

\begin{frontmatter}


\title{Characterization of electron density and ionization of a uranium laser produced plasma using laser absorption spectroscopy}

\author[label1] {Ryland\,G.\,Wala}
\author[label2,label3] {Mathew\, P.\ Polek}
\author[label3] {Sivanandan\,S.\,Harilal}
\author[label1] {R.\,Jason\,Jones}
\author[label1] {Mark\,C.\,Phillips}

\cortext[correspondingauthor1]{Corresponding author: mphillips@optics.arizona.edu}

\address[label1]{James C. Wyant College of Optical Sciences, University of Arizona, Tucson, Arizona 85721, USA }
\address[label2]{Center for Energy Research, University of California San Diego, La Jolla, California 92093, USA}
\address[label3]{Pacific Northwest National Laboratory, Richland, Washington 99352, USA}

\begin{abstract}
High-resolution tunable laser spectroscopy is used to measure time-resolved absorption spectra for ten neutral uranium transitions and six singly-ionized transitions in a laser produced plasma. Spectral lineshapes are analyzed to determine temporal variations in ion and neutral total column densities, excitation temperatures, kinetic temperatures, and collisional broadening effects as the plasma cools. Comparison of ion to neutral column densities shows a ratio greater than 10 at times $<$15~$\mu$s after plasma onset, with the ratio not reaching unity until $\sim$50~$\mu$s. Spectral lineshapes are analyzed to separate Stark and van der Waals contributions to collisional broadening, from which electron densities are determined and found to decrease from $\sim$10$^{15}$-10$^{13}~$cm$^{-3}$ over times from 4-25~$\mu$s. Using absorption spectroscopy to determine charge properties and electron density over these time scales and at low magnitudes provides valuable insight into plasma properties not obtainable using conventional emission spectroscopy. Comparisons between ion and neutral densities, excitation temperatures, kinetic temperatures, and electron densities could indicate potential deviations from local thermodynamic equilibrium and Saha ionization predictions.
\end{abstract}

\begin{keyword}
Laser absorption spectroscopy \sep laser-induced breakdown spectroscopy (LIBS)  \sep spectral modeling  \sep uranium \sep laser produced plasma
\end{keyword}

\end{frontmatter}

\date{\today}

\section{Introduction}
\label{Introduction}

Optical spectroscopy of laser produced plasmas (LPPs) provides a powerful tool for non-contact elemental analysis of solid samples, and studying fundamental properties of atoms and molecules in plasma conditions \cite{2022-Hari-RMP, majidi1992spectroscopic}. Measurements of uranium in LPPs are of high interest as a method to assist with elemental/isotopic analysis of solid materials used as nuclear fuels, found as debris, or present as by-products of energy/weapons production processes \cite{HariAPR2018}. Extensive research has been devoted toward development of analytical methods for uranium elemental and isotopic analysis using Laser Induced Breakdown Spectroscopy (LIBS) \cite{GeorgeChan2013, GeorgeChan2016-U-soil-isotopes, Cremers2012, Doucet2011, Pietsch1998, Russo2024ULIBS}, and also laser absorption and fluorescence-based methods \cite{PhillipsSR2017, Miyabe2013, Miyabe2012Ce, hull2022combined, Liu2002LAS, Quentmeier2001,smith1999measurement}. Uranium LPPs can also be used as laboratory-scale surrogates of nuclear fireball conditions where there is interest in understanding the conditions for nuclear fallout formation; specifically, the physical/chemical pathways of uranium from initial plasma conditions, to molecular species such as U$_x$O$_y$, and eventually to solid particulate material \cite{Kautz_2021_UReview}.

In all these applications, understanding the physical properties of LPPs, including ionization states, excitation temperature, and electron density, are critical for predicting the chemical and optical properties of uranium and other species, and these properties must be known throughout the evolution of the LPP \cite{Kautz_2021_UReview}. Experimental studies on fundamental properties of uranium in LPPs have been used to measure excitation \cite{SkrodzkiSCAB2016, SinghPST2018, HarilalPCCP2019, BurgerPoP2019, Weerakkody2021, WeerakkodySCAB2020, polek2024comparison} and kinetic \cite{phillips2023, taylor2014differential} temperatures, measure electron density at early stages of LPP evolution \cite{SinghPST2018, hull2022combined, BurgerPoP2019, SkrodzkiSCAB2016}, determine van der Waals \cite{taylor2014differential} and Stark broadening coefficients \cite{BurgerPoP2019}, measure hyperfine constants \cite{HarilalSCAB2020}, and study conditions for uranium oxide molecular formation \cite{Weerakkody2021, WeerakkodySCAB2020, 2019-JAAS-Liz, HartigOE2017, SkrodzkiOL2018, SkrodzkiSCAB2016, SkrodzkiPoP2019, Milos2021-review, Weisz2017, burton2022effect}. Despite these prior studies, there is limited information on the late-time dynamics of uranium LPPs, including lifetimes of ionic species and electron densities under different experimental conditions. 

Most often, properties of LPPs are characterized by optical emission spectroscopy (OES), or LIBS~\cite{2022-Hari-RMP,SinghLIBSbook}. These emission methods provide valuable information on the early-time properties of LPPs, including atomic/ionic excitation temperatures and electron density. Because these methods require a luminous plasma, they are most appropriate for probing early times of LPP evolution when temperatures and emission intensities are high, and become challenging for times $\gtrsim$10~$\mu$s. Excitation temperatures are retrieved by comparing the relative emission intensity of species in the same or different charge states that are assumed to be in local thermodynamic equilibrium (LTE) using Boltzmann or Saha-Boltzmann analysis~\cite{2022-Hari-RMP}. The electron density is typically retrieved from the Stark broadened linewidth of a spectral feature~\cite{2022-Hari-RMP,Konjevic_specdiagLIP}. However, as the LPP cools, the Stark broadening becomes smaller than the instrumental lineshape, and emission intensities become small, making the determination of the electron density challenging. Strong emission lines may also be subject to self-absorption effects, which can limit the ability to obtain accurate lineshape information \cite{Konjevic_specdiagLIP}. In some cases, Saha-Boltzmann analysis may be used to determine electron density; however, uncertainties are typically higher than Stark-broadening methods \cite{Sarkar2017_SB}.

Absorption spectroscopy may be used for probing LPPs at later time delays than OES, because it measures the population in lower energy levels that remain populated at lower temperatures \cite{2022-Hari-RMP}. Absorption-based methods readily obtain quantitative state column densities of atomic and molecular species as long as transitions can be measured above noise levels, and if oscillator strengths are known \cite{MERTEN2022review, 2022-Hari-RMP, Weerakkody2021,Kane_Fe}. Absorption spectroscopy of LPPs has been performed using a range of light sources, including flashlamp discharges \cite{Weerakkody2021, WeerakkodySCAB2020}, broadband
emission from a LPP \cite{ribiere2010analysis,nagli2012fraunhofer, costello1991xuv}, optical parametric oscillators (OPOs) \cite{merten2022laser, merten2023following}, frequency combs \cite{Bergevin2018, Kane_Fe, mccauley2024dual, 2021-SCAB-Weeks, rhoades2022dual}, and tunable continuous wave lasers \cite{PhillipsSR2017, Miyabe2013, Miyabe2012Ce, hull2022combined, Liu2002LAS, Quentmeier2001,smith1999measurement, taylor2014differential}. Similar to OES methods, when multiple absorbing transitions of a given species are measured, Boltzmann analysis may be used to obtain quantitative total column/number densities \cite{2022-Hari-RMP}. Multiple ionization stages of elements in LPPs have been measured using absorption spectroscopy \cite{WeerakkodySCAB2020, Weerakkody2021, merten2023time, mccauley2024dual}, and Saha-Boltzmann analysis used to extract excitation temperature and electron density \cite{mccauley2024dual}.

While absorption methods should be able to measure Stark broadening and associated electron densities in LPPs at later times than OES, there are challenges associated with this approach. Broadband absorption spectroscopy methods may not have sufficient spectral resolution to resolve small variations in spectral linewidth/lineshape required for resolving weak Stark broadening at low electron densities. At early times in LPP evolution when Stark widths are large, high excitation temperatures may depopulate lower energy levels, leading to low absorption strengths and poor signal-to-noise ratio (SNR), also inhibiting measurement of Stark widths until the LPP cools. As Stark broadening becomes smaller, differentiating it from other broadening mechanisms including Doppler and van der Waals \cite{Gornushkin1999, 2022-Hari-RMP} becomes more challenging, and extraction of these parameters from Voigt lineshape fits requires high SNR spectra. Furthermore, interpretation of Voigt fit parameters requires assumptions about plasma spatial uniformity, which may break down at early times of LPP evolution. For example, in low pressure LPPs, Doppler splitting may cause the observed absorption lineshape to deviate strongly from a Voigt profile \cite{Miyabe2012Ce,bushaw1998investigation}. Tunable laser absorption spectroscopy (LAS) has sufficient spectral resolution and high SNR to obtain meaningful Voigt fits at late times of LPP evolution, and has been used to determine kinetic temperatures from Gaussian widths \cite{phillips2023, taylor2014differential,2021-SCAB-Lahaye, 2021-PRE-Hari}, and van der Waals broadening from Lorentzian widths \cite{taylor2014differential, 2021-PRE-Hari}. Building upon these prior studies and extending the analysis of absorption lineshapes to measure Stark broadening will allow determination of lower magnitudes of electron density and over time scales $\gtrsim$10~$\mu$s, providing new information on LPP charge properties difficult to measure with OES of LPP (LIBS).

In this paper, we report high-resolution and time-resolved measurements of ten neutral and six singly-ionized transitions of uranium atoms in a LPP in 50~Torr argon. Spectral fitting of these transitions is used to extract total column densities of both neutral (U~I) and singly-ionized (U~II) uranium, along with excitation temperatures. The results show that uranium ionic absorption is detected in the LPP for $>$100~$\mu$s after plasma formation, while absorption from neutral transitions is observed for $>$1,000~$\mu$s. The ionization ratio U~II / U~I is greater than 10 for times $<$15~$\mu$s, and remains greater than unity until $\sim$50~$\mu$s, showing a high degree of ionization and persistence of ions in the uranium LPP. Analysis of absorption lineshapes for U I and U II transitions is used to separate contributions from van der Waals and Stark broadening, and to estimate the electron density from 4-25~$\mu$s, yielding values ranging from $\sim10^{15}-10^{13}~$cm$^{-3}$. Despite the U~I and U~II state populations each following Boltzmann distributions, differences are observed between ion and neutral excitation temperatures and kinetic temperatures, and strong deviations from Saha equilibrium predictions are apparent, indicating possible departures from LTE or the presence of long-lived spatial inhomogeneity of the LPP.

\section{Experimental Details}
\label{methods}

The experimental setup used for the present study was identical to that of previous measurements \cite{phillips2023, Hari-JAP-2022, HariPSST2021}. The LPPs were generated using 55 mJ pulses from a 1064~nm Nd:YAG laser with a 6~ns pulse width, repetition rate of 10~Hz, and a focused beam diameter of $\sim$0.65~mm. The sample was uranium metal with natural isotopic abundance, and was held in a vacuum chamber under flowing Ar at 50~Torr pressure. The sample was cleaned of the oxide layer using several ablation shots before each measurement. A continuous-wave (CW) tunable Ti:Sapphire laser with $\sim$100~kHz linewidth was used to measure time-resolved absorbance spectra from selected U~I and U~II transitions in a wavelength region from 763-835~nm. The probe laser had a power of 3~mW, a diameter of $\sim$0.7~mm, and passed through the LPP at a distance of $\sim$1~mm above the sample surface. The transmitted probe intensity was detected with a Si photodiode with a 200~ns rise time and digitized at 2~MHz. The probe wavelength was stepped across the targeted transition, and the time-resolved transmission was measured for eight ablation shots at each probe wavelength. The measured time-resolved intensity $I(\lambda,t)$ was converted to absorbance $A(\lambda,t)$ using the average pre-ablation probe intensity $I_0(\lambda)$ via $A(\lambda,t)=-\ln[I(\lambda,t)/I_0(\lambda)]$, and the absorbance from the multiple ablation shots at each wavelength was averaged. The wavelength/frequency at each step was recorded using a wavemeter. The resulting dataset provides a map of absorbance versus wavelength and time. 

For the experiments reported here, ten U~I optical transitions ranging from 787-835~nm, and six U~II optical transitions ranging from 763-808~nm were measured. The line parameters for these transitions are listed in Table \ref{Used Lines}, which were obtained from the Palmer atlas \cite{PalmerAtlas1980}. Within the accessible Ti:Sapphire laser tuning range the Palmer atlas lists 10 U~II lines, 325 U~I lines, and 34 unassigned lines. A subset of these lines was selected primarily based on oscillator strength and lower energy level, to provide a range of transitions visible at early and late times of plasma evolution, and for use in Boltzmann analysis. Some potentially desirable lines were rejected based on partially-overlapping (blended) lines that could not be accounted for in the spectral fits. For example, within this spectral region there are 6,218 U~II and 8,817 U~I possible dipole-allowed transitions based on the uranium energy levels \cite{Blaise1976, Blaise1994}, most of which are not listed in Palmer \cite{PalmerAtlas1980}. Absorption from the $\mathrm{^{235}U}$ minor isotope with 0.7$\%$ abundance was neglected as it was not typically visible above the noise level in the fitted spectra, and would be a minor correction relative to other experimental uncertainties.

\section{Results}
\label{results_discussion}

\subsection{Measurement and Fitting of Absorbance Spectra}

\begin{table*}
\caption{\label{Used Lines} 
Spectroscopic parameters of U~I and U~II transitions used in these experiments, obtained from Palmer et al. \cite{PalmerAtlas1980}. $\lambda_{air}$ (nm) \& $\nu$ (GHz) are the wavelength (in air) and transition frequency, E$_i$ (cm$^{-1})$ and E$_j$( cm$^{-1})$ denote the lower (i) and upper (j) energy levels, J$_i$ and J$_j$ denote the respective total angular momentum quantum numbers, and $f_{ij}$ is the transition oscillator strength calculated according to Eqn. 7 in Ref. \cite{PalmerAtlas1980}.
}

\begin{center}
\small

\begin{tabular}{|c|c|c|c|c|c|c|c|} 

\hline
Charge State & $\nu (GHz)$ & $\lambda_{air}$ (nm) & E$_i$ (cm$^{-1})$ & J$_i$ & E$_j$( cm$^{-1})$ & J$_j$ & $f_{ij}$ \\ 
\hline

I & 380886.9 & 786.8740 & 7864 & 5 & 20569 & 4 & 2.8$\times$10$^{-3}$ \\ 
I & 380884.6 & 786.8788 & 3800 & 7 & 16505 & 6 & 1.9$\times$10$^{-3}$ \\ 
I & 380565.1 & 787.5393 & 11308 & 9 & 24002 & 8 & 1.4$\times$10$^{-2}$ \\
I & 379740.2 & 789.2501 & 20218 & 5 & 32885 & 6 & 1.9$\times$10$^{-2}$ \\
I & 379038.9 & 790.7103 & 0 & 6 & 12643 & 6 & 8.7$\times$10$^{-5}$ \\
I & 378478.4 & 791.8815 & 4275 & 6 & 16900 & 7 & 2.2$\times$10$^{-3}$ \\
I & 376658.6 & 795.7073 & 10069 & 7 & 22633 & 7 & 1.4$\times$10$^{-3}$ \\
I & 376332.8 & 796.3963 & 10081 & 5 & 22634 & 4 & 2.0$\times$10$^{-2}$ \\
I & 376026.1 & 797.0458 & 8118 & 7 & 20661 & 6 & 2.3$\times$10$^{-2}$ \\
I & 359073.7 & 834.6758 & 5991 & 4 & 17968 & 3 & 6.1$\times$10$^{-3}$ \\ 
\hline 
II & 392659.7 & 763.2816 & 2294 & 11/2 & 15392 & 13/2 & 2.2$\times$10$^{-5}$ \\
II & 385201.2 & 778.0609 & 4585 & 13/2 & 17434 & 11/2 & 6.0$\times$10$^{-5}$ \\
II & 384124.6 & 780.2416 & 10740 & 11/2 & 23553 & 11/2 & 4.9$\times$10$^{-4}$ \\
II & 382394.6 & 783.7716 & 9075 & 7/2 & 21831 & 9/2 & 5.0$\times$10$^{-4}$ \\
II & 375722.7 & 797.6893 & 5667 & 7/2 & 18200 & 9/2 & 2.8$\times$10$^{-4}$ \\
II & 371202.2 & 807.4037 & 6445 & 9/2 & 18827 & 11/2 & 2.1$\times$10$^{-4}$ \\  
\hline

\end{tabular}
\end{center}
\end{table*}

 \begin{figure}
\centering
\includegraphics[width=0.45\textwidth]{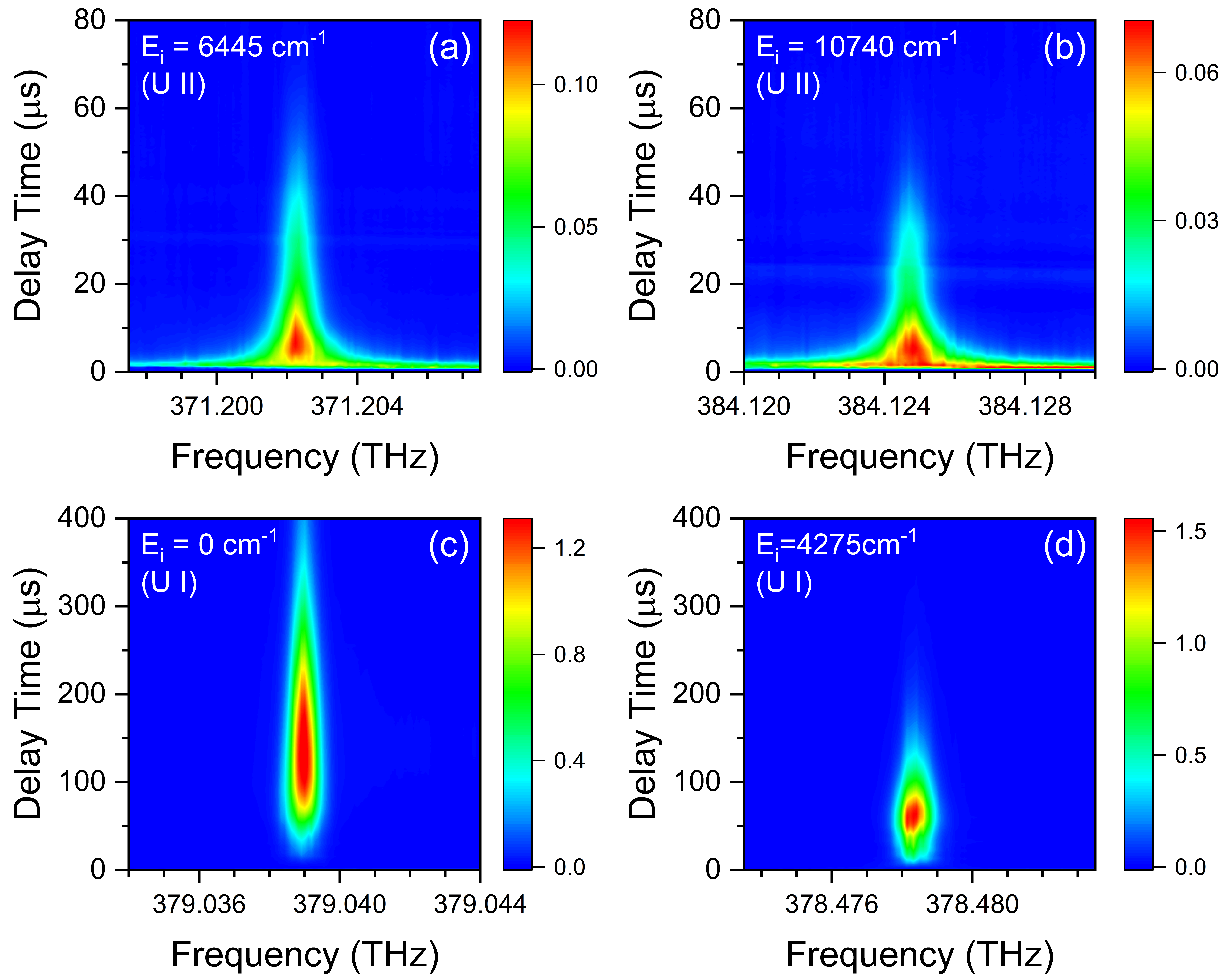}
\caption{\small{Measured time-resolved absorbance spectra of U~II transitions with lower energy levels of (a) 6,445~cm$^{-1}$ and (b) 10,740~cm$^{-1}$ and U~I transitions with lower energy levels of (c) 0~cm$^{-1}$ and (d) 4,275~cm$^{-1}$.}}
\label{absorption spectra}
\end{figure}

 \begin{figure}
\centering
\includegraphics[width=0.45\textwidth]{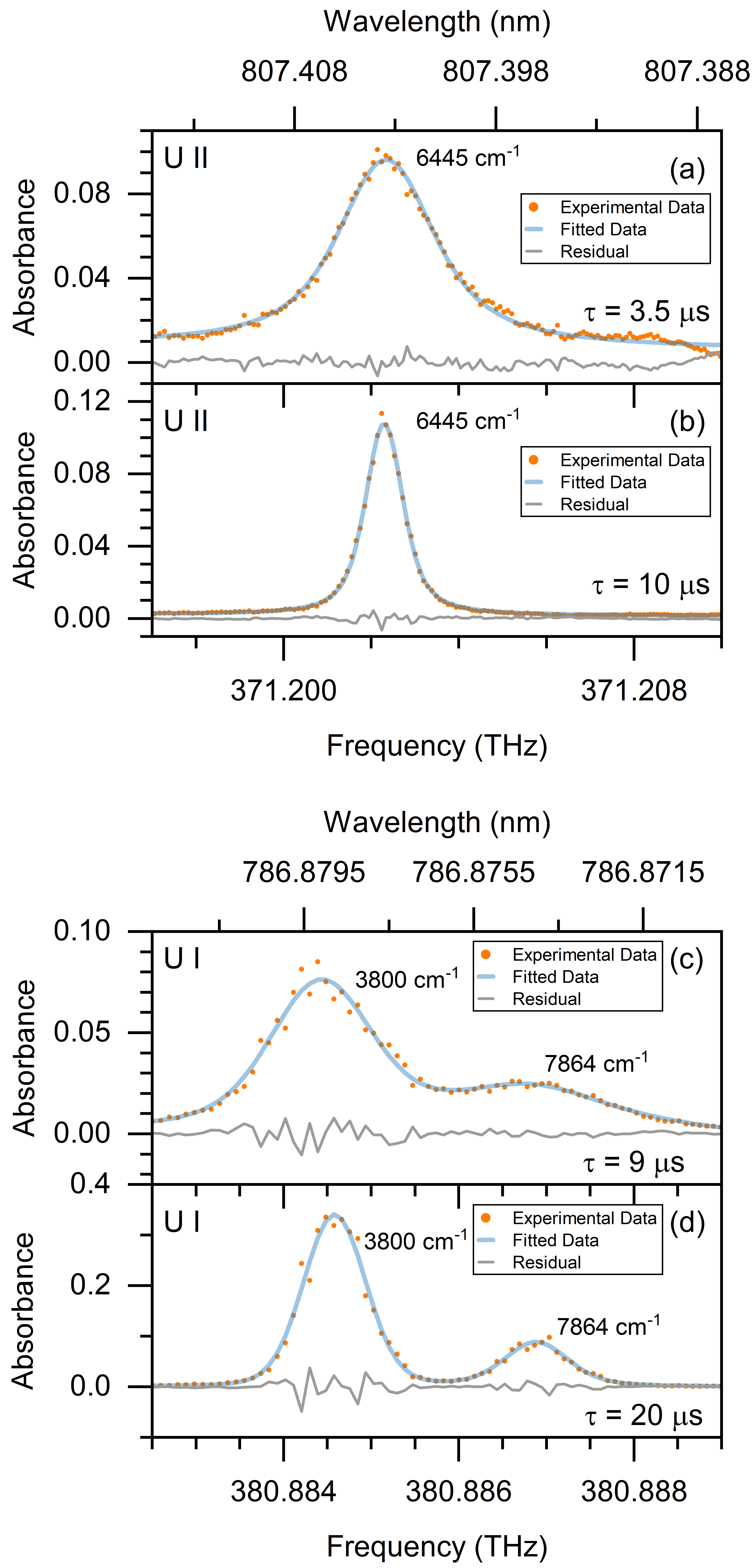}
\caption{\small{Experimental absorbance spectra (solid points), spectral fits (blue line), and fit residuals (gray line). The top panels show a single U~II transition with a lower energy level of 6,445~cm$^{-1}$ at (a) 3.5~$\mu$s and (b) 10~$\mu$s. The lower panels show two U~I transitions with lower energy levels of 3,800~cm$^{-1}$, and 7,864~cm$^{-1}$ at (c) 9~$\mu$s and (d) 20~$\mu$s.}}
\label{Fits}
\end{figure}

Figure~\ref{absorption spectra} shows examples of experimental absorbance data. The magnitude of the absorbance is given by the color mapping, while the horizontal and vertical axes show the optical frequency in THz and the time after ablation in microseconds. Figure~\ref{absorption spectra}(a,b) show time-resolved absorbance spectra for two measured U~II transitions with lower energy levels of 6,445 (807.4037~nm) and 10,740~cm$^{-1}$ (780.2416~nm), respectively. Figure~\ref{absorption spectra}(c,d) show time-resolved absorbance spectra for two measured U~I transitions with lower energy levels of 0 (790.7103~nm) and 4,275~cm$^{-1}$ (791.8815~nm), respectively. The U~II transitions shown in Fig.~\ref{absorption spectra}(a,b) are stronger at early times of the plasma evolution and show an initial large spectral broadening which decreases with time. The U~I transitions shown in Fig.~\ref{absorption spectra}(c,d) initially have lower absorbance and reach a maximum at later times in LPP evolution, before also decaying at late times. The U~II transitions persist for shorter times relative to the U~I transitions. Similarly, transitions originating from higher energy levels persist for shorter times relative to those originating from low energy levels, with the ground state U~I transition having the longest persistence. 

For each measured transition, the absorbance spectrum at each 0.5~$\mu$s time step was fit using a model composed of a sum of Voigt profiles including the target transition and any visible secondary (interfering) peaks, along with a linear baseline \cite{2021-SCAB-Weeks, phillips2023}. For each transition, the output of the fitting procedure is a time-series of peak center frequencies, peak areas, Lorentzian linewidths, Gaussian linewidths, and Voigt linewidths. Figure~\ref{Fits} shows examples of spectral fits to experimental data, along with fit residuals. Figures~\ref{Fits}(a) and \ref{Fits}(b) show spectral fits for a U~II transition originating from a lower energy level of 6,445~cm$^{-1}$ at 3.5 and 10~$\mu$s, respectively, highlighting the decrease in peak width with time. Figures~\ref{Fits}(c) and \ref{Fits}(d) show spectral fits at 9 and 20~$\mu$s, respectively, for a single scan region that contains two U~I transitions. In this case, the leftmost peak originates from an energy level of 3,800~cm$^{-1}$ and the rightmost peak from an energy level of 7,864~cm$^{-1}$. Despite the small peak separation of 2.3~GHz (4.8~pm), the two transitions are clearly resolved in the experimental spectrum and fit. Similar to the U~II transitions, the U~I widths decrease with time.

Figure~\ref{Fits} also shows the fit residuals in gray. Near the peak centers, the residuals show higher noise due to variations in ablation conditions between spectral points (flicker noise), but there is no obvious structure in the residuals that would indicate a bad spectral fit. Spectral fits with poor residuals (high structure) were observed for some transitions, especially for U~I transitions at early times $\lesssim$~10~$\mu$s with low peak absorption, and fit results at these times were excluded from further analysis. In other cases, poor fits were identified as arising from closely-separated overlapping peaks not included in the fitting model, and these results were also discarded from further analysis. For some spectral fits with residuals appearing visually to be good, the Lorentzian and Gaussian widths yielded non-physical results (e.g. zero) due to low SNR or possibly due to deviations from a Voigt profile in the measured peak shape. For these cases, the peak area was used in Boltzmann analysis; however, the Gaussian and Lorentzian widths were not used in lineshape analysis. U I transitions at 786.8740, 787.5393, 789.2501, and 795.7073~nm were not used in lineshape analysis for this reason. 

\begin{figure}
\centering
\includegraphics[width=0.45\textwidth]{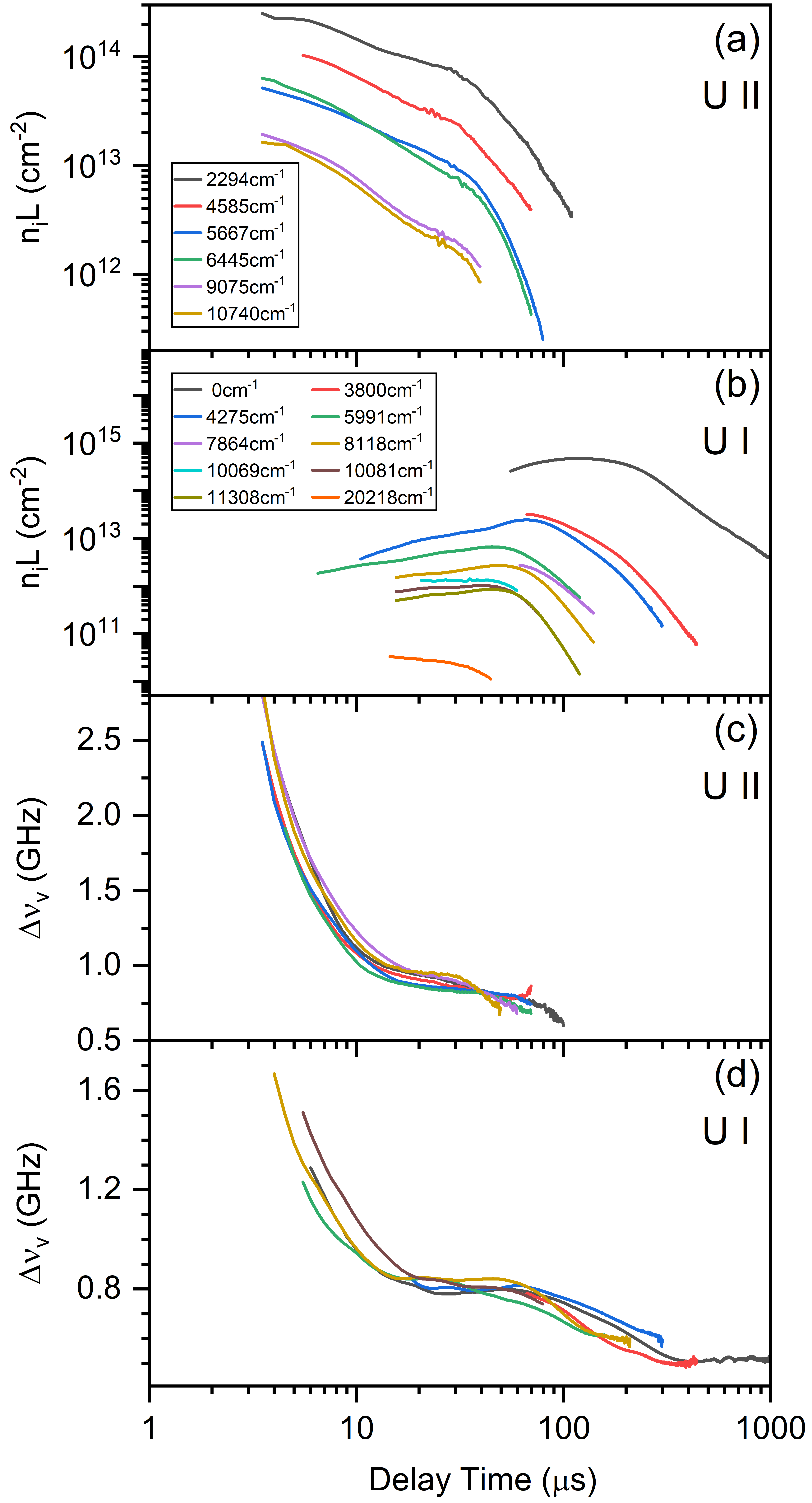}
\caption{\small{Measured state column densities for (a) ions and (b) neutrals as a function of time. Measured Voigt linewidths used for linewidth analysis for (c) ions and (d) neutrals versus time.}}
\label{StateVoigt}
\end{figure}

The state column density $n_iL(t)$ for each transition was calculated from the fitted peak area using $n_iL(t)=\frac{Area(t)}{\tilde{\sigma_0} f_{ij}}$ where $f_{ij}$ is the the transition oscillator strength and $\tilde{\sigma_0}=\frac {e^2}{4 \epsilon_0 m_e c}$ is the integrated absorption cross-section constant \cite{2022-Hari-RMP}. State column densities for the measured U~I and U~II transitions are shown in Figure~\ref{StateVoigt}(a,b). The results show expected trends according to Boltzmann distributions, with lower $E_{i}$ states being more populated and longer lived relative to higher energy levels of the same charge state. It is also apparent that the U~II transitions are consistently visible at earlier time delays, while many U~I transitions could not be detected and fit until after 10~$\mu$s. Figure~\ref{StateVoigt}(c,d) show the Voigt full-width at half maximum (FWHM) for the measured U~I and U~II peaks. The results show good consistency between the different transitions in the same charge state, with all following the same trends with time. The Voigt widths decrease with time as expected due to decreasing temperature and electron density, ranging $\sim$2.5~GHz (5.2~pm) for U~II at 3.5~$\mu$s to $\sim$0.5~GHz for U~I at $t>$~400~$\mu$s. 

\subsection{Boltzmann Analysis}

\begin{figure}
\centering
\includegraphics[width=0.45\textwidth]{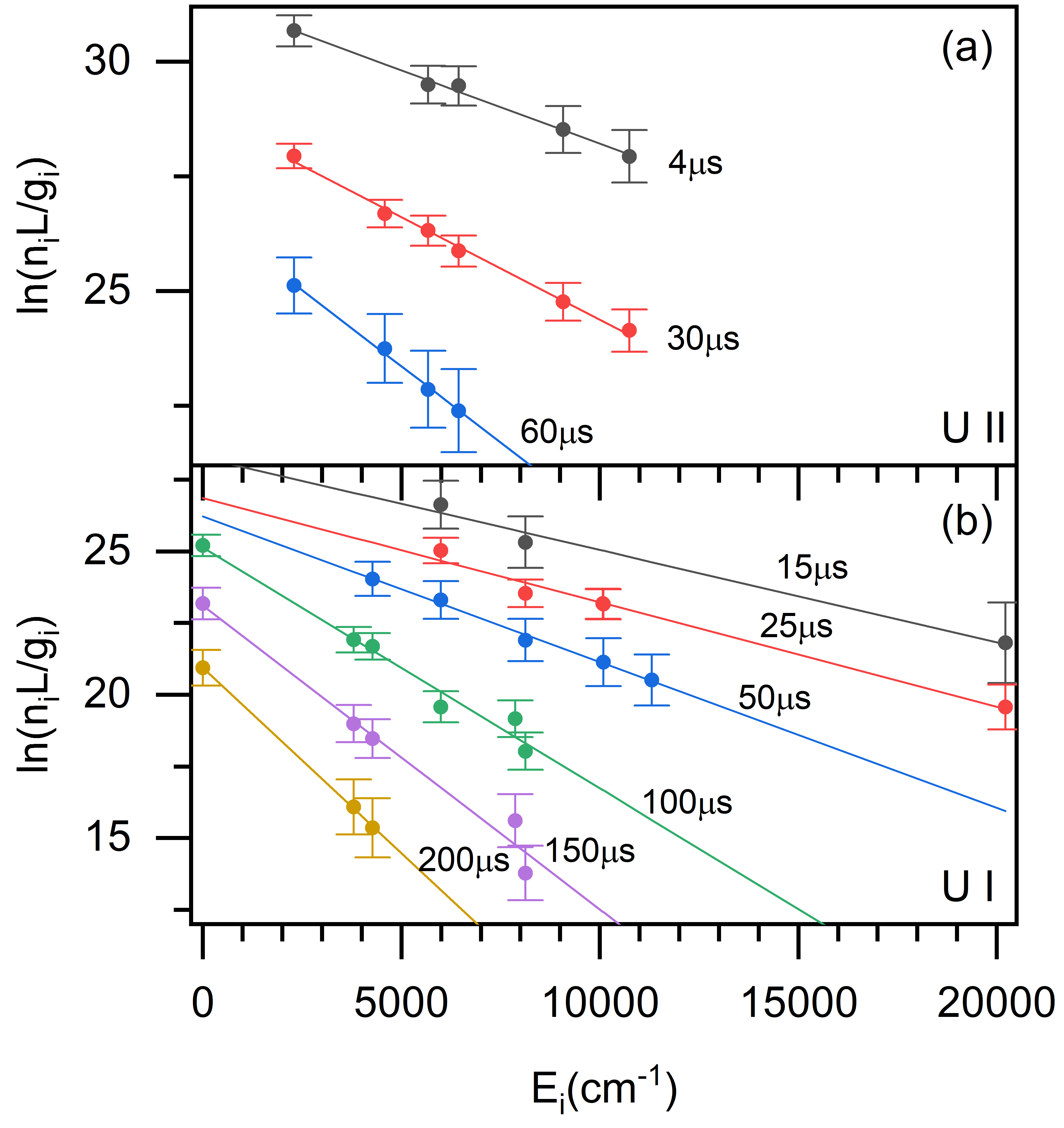}
\caption{\small{Boltzmann plots for selected times indicated in figure. Solid points are experimental data, and solid lines are linear fits. Plots are offset for clarity. (a) U~II transitions. (b) U~I transitions}.}
\label{Boltzmann}
\end{figure}

The state column densities for the ten U~I transitions and six U~II transitions were used to construct Boltzmann plots for U~I and U~II at each time delay, and examples are shown in Figure~\ref{Boltzmann}. Note that the number of points in each Boltzmann plot changes with time according to the number of transitions which could be measured and fit at each time step. Error bars are shown based on a 20\% uncertainty in the oscillator strengths \cite{PalmerAtlas1980}. The y-intercept of the Boltzmann plot was used to determine the total column density of the species and the slope was used to determine the excitation temperature \cite{2022-Hari-RMP}. Calculation of the total column density requires the partition function, which was calculated from U~I and U~II energy level data \cite{Blaise1976,Blaise1994}. The Boltzmann plots display good linearity, indicating that each charge state is well described by a single temperature and the level populations follow a Boltzmann distribution.

\begin{figure}
\centering
\includegraphics[width=0.45\textwidth]{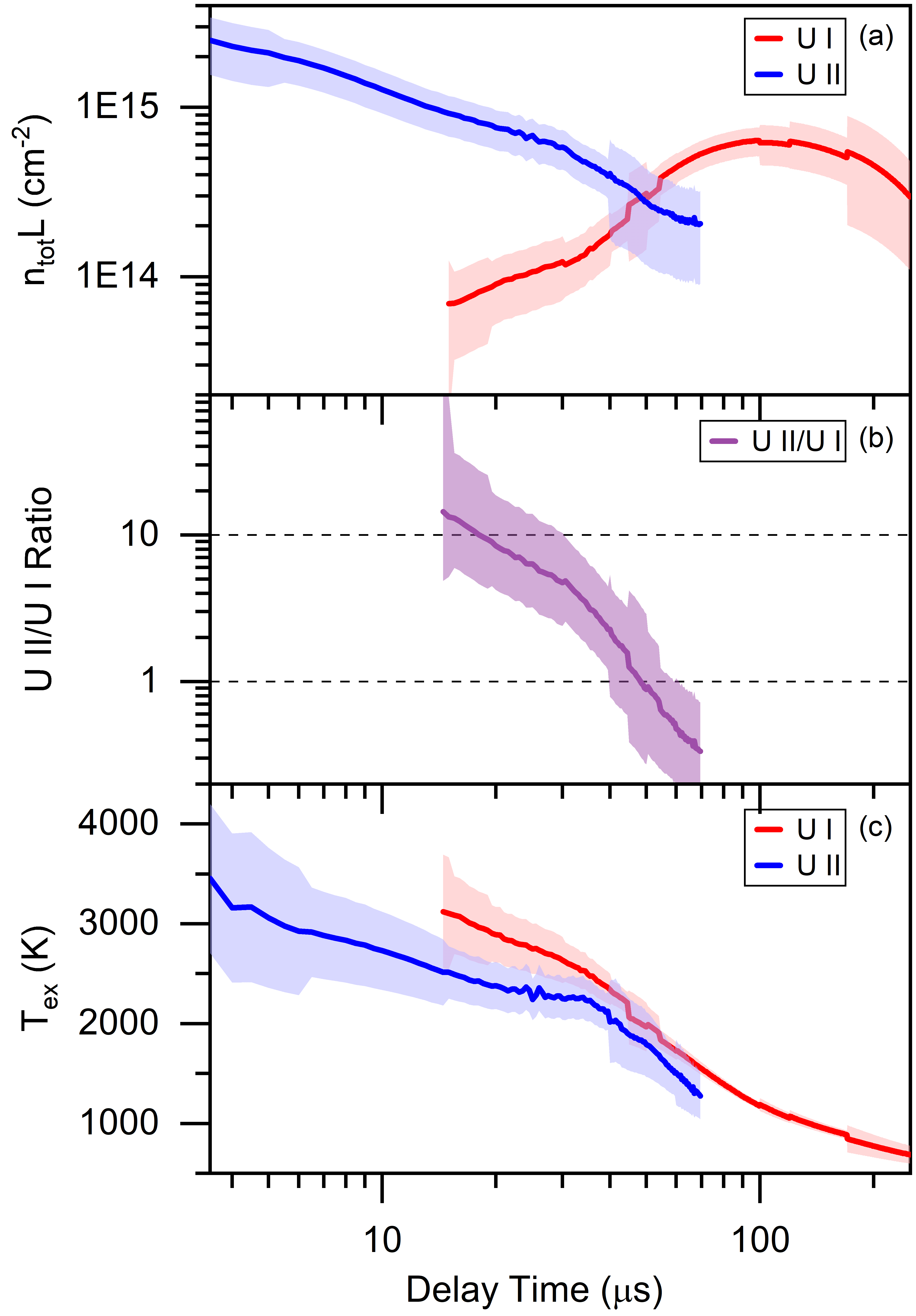}
\caption{\small{(a) Total column density for U~I (red) and U~II (blue) as a function of time. The shaded region shows the estimated uncertainty. (b) The ratio of U~II to U~I total column densities as a function of time. (c) Excitation temperature versus time for U~I (red) and U~II (blue).}}
\label{Density and Temp}
\end{figure}

The total column densities for the ions and neutrals are shown in Figure~\ref{Density and Temp}(a). The surrounding shaded region shows the estimated uncertainty calculated from the standard error of the linear Boltzmann fit at each time. The total column density accounts for U~I and U~II in all energy levels, and the results show that at early delay times the plasma is highly ionized. As the LPP expands and cools, the U~II column density decreases, while the U~I column density increases, due to electron-ion recombination. The U~I column density reaches a maximum near 100~$\mu$s, and is consistent with previous results obtained under similar ablation conditions \cite{phillips2023}.
The ratio of U~II to U~I column density is shown in Figure~\ref{Density and Temp}(b). The results show a 14.42 ratio of the ion to neutral column densities at $\sim$15~$\mu$s, which is the earliest time for which the U~I and U~II column densities could both be determined. The ratio of ion to neutral column densities does not reach one until $\sim$50~$\mu$s. At later delay times, the ion to neutral ratio could not be determined from Boltzmann analysis due to the low absorption of the ionic transitions probed. 

The excitation temperatures determined from the Boltzmann plot fits are displayed in Figure~\ref{Density and Temp}(c). The U~II excitation temperature is determined at earlier times than the neutrals due to the higher absorption signal and corresponding better spectral fits at these early times. The ion excitation temperature could be determined from 3.5~$\mu$s to $\sim$70~$\mu$s, while the neutral excitation temperature was determined from 15~$\mu$s to $\sim$250~$\mu$s. The measured ionic excitation temperature appears to be lower than the neutral excitation temperature, which is not consistent with prior results typically showing a higher ion than neutral excitation temperature \cite{WeerakkodySCAB2020,2021-SCAB-Lahaye,mccauley2024dual}. A higher observed ion versus neutral excitation temperature is usually attributed to differences in spatial distributions between the two charge states \cite{2022-Hari-RMP, Aguilera2004}. In the current experiments, the probe beam was 1~mm above the sample surface, which could affect the spatial distributions of ion and neutral atoms along the measurement path at different times \cite{Hari-JAP-2022} and lead to different observed temperatures. However, in the overlap region where ion and neutral excitation temperatures were both determined, the values are equal within the uncertainty ranges and so additional experiments would be required to determine if the temperature differences are physically significant.

\subsection{Lineshape Analysis}

\begin{figure}[]
\centering
\includegraphics[width=0.45\textwidth]{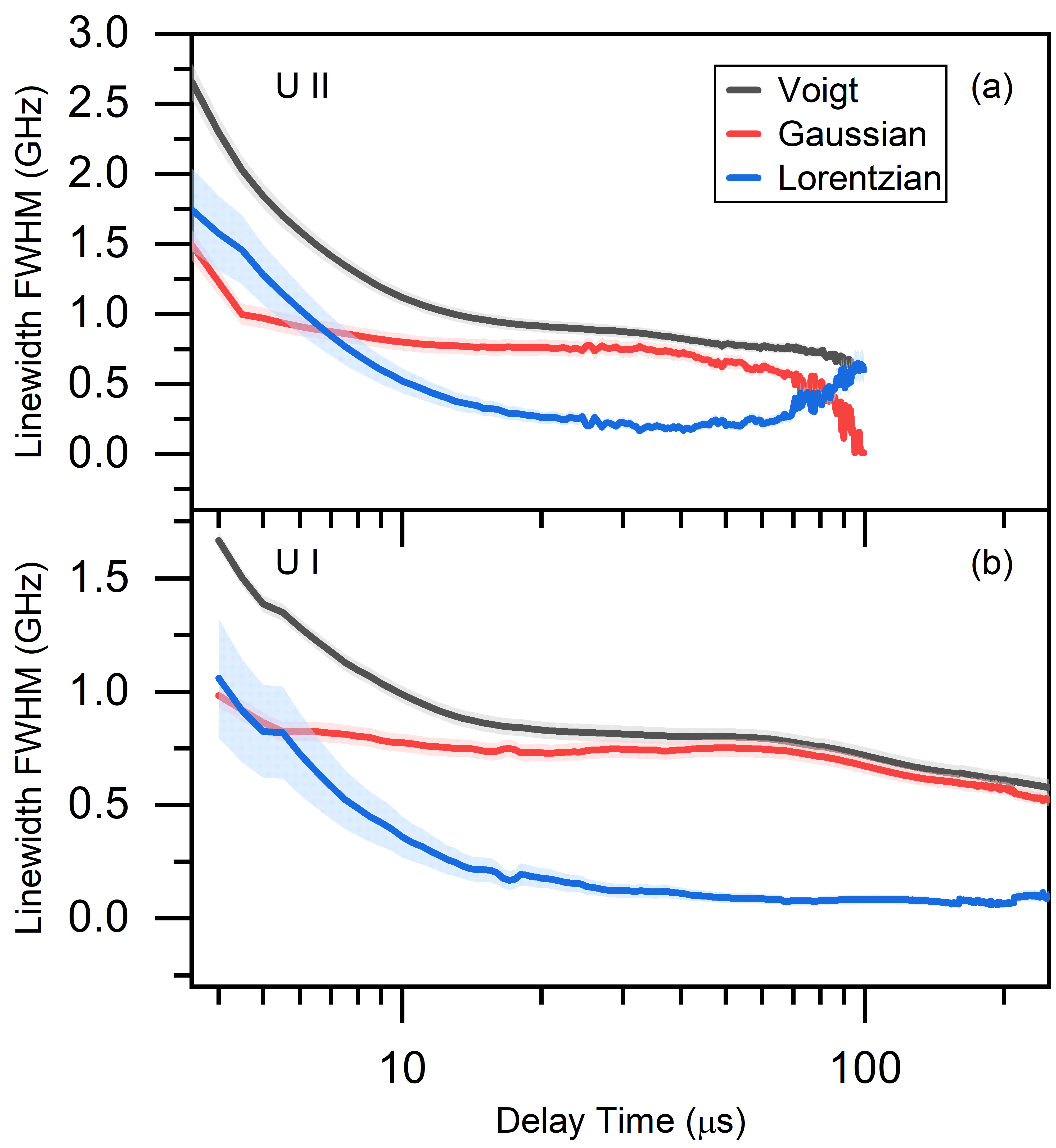}
\caption{\small{Mean measured Voigt (black), Gaussian (red), and Lorentzian (blue) linewidths for (a) U~II and (b) U~I. The shaded regions show uncertainties derived from the standard deviations over the six measured transitions. 
}}
\label{Voigt Gaussian Lorentzian}
\end{figure}

As was shown in Fig.~\ref{StateVoigt}(c,d), the Voigt FWHM from spectral fits showed similar values between the various U I transitions, and between the various U II transitions, although the values for U I were different from U II. The Gaussian/Lorentzian linewidths showed a similar behavior. The solid lines in Fig.~\ref{Voigt Gaussian Lorentzian} show the mean Voigt, Lorentzian, and Gaussian FWHM versus time for the six U~I and six U~II transitions which were analyzed for lineshapes. At times for which all six transitions could be measured, the relative standard deviations of (U~II,~U~I) linewidths were (5\%,~4\%) for the Voigt FWHM, (8\%,~5\%) for the Gaussian FWHM, and (17\%,~25\%) for the Lorentzian FWHM, and these uncertainties are shown by the shaded bands. The higher uncertainty for the Lorentzian FWHM results because it is a smaller fraction of the total linewidth for most times. The results in Fig.~\ref{Voigt Gaussian Lorentzian} indicate that the Gaussian and Lorentzian components of the total Voigt width change differently with time, which reflects the different physical mechanisms behind each component.

The Gaussian component of the linewidth arises from Doppler broadening, and was used to calculate the kinetic (Doppler) temperature $T_{k}$ using:

\begin{align}
\centering
    \Delta\lambda_{g}=7.16\times10^{-7} \lambda_{0}\sqrt{\frac{T_{k}}{m}}
\label{Doppler}
\end{align}

\begin{figure}
\centering
\includegraphics[width=0.45\textwidth]{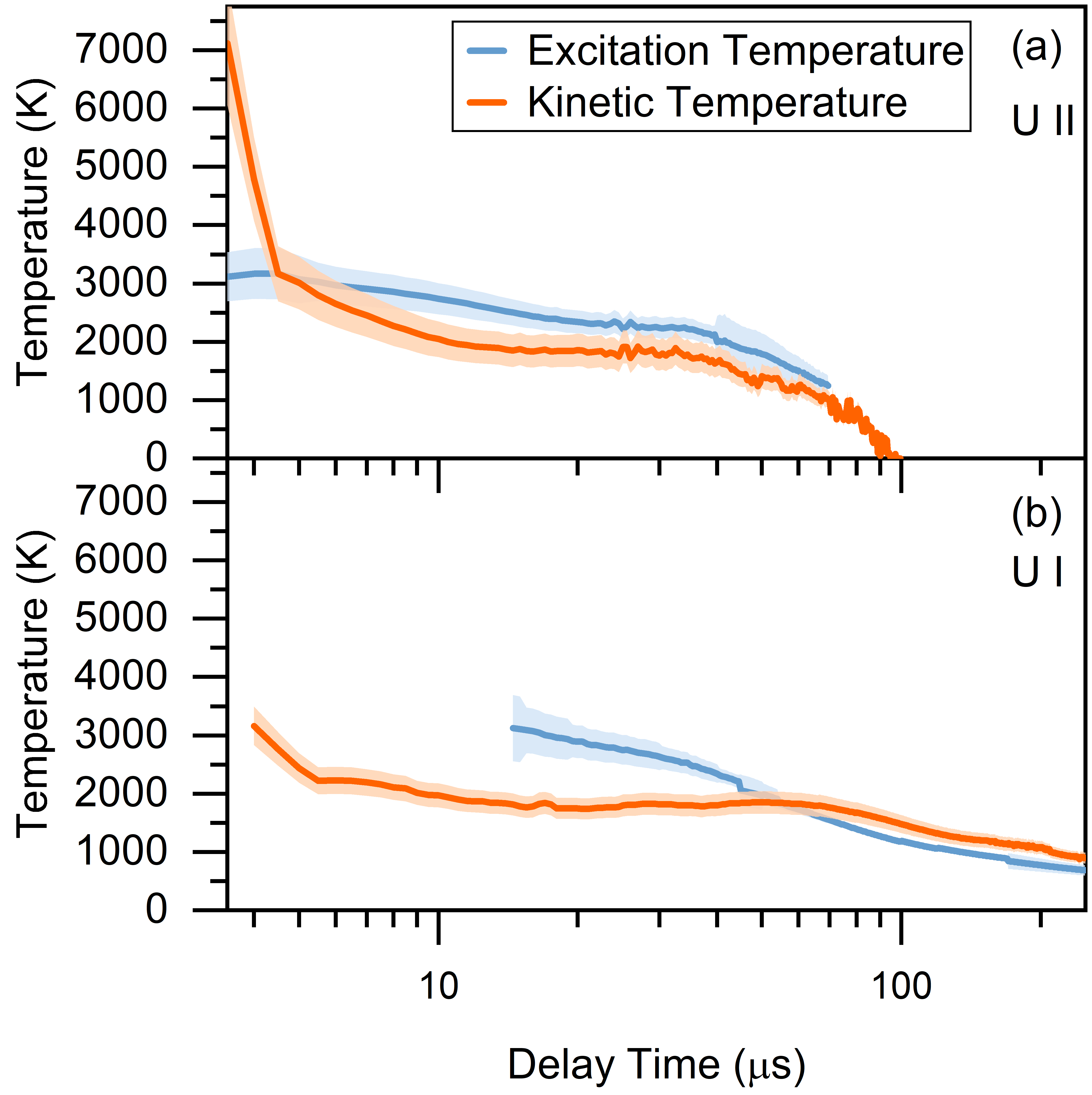}
\caption{\small{Excitation temperature from Boltzmann analysis (blue) and kinetic temperature from average Gaussian linewidth (orange) for (a) U~II transitions and (b) U~I transitions, as a function of time.}}
\label{Temperatures}
\end{figure}

Where $\Delta\lambda_{g}$ is the Gaussian full width at half max (FWHM), $\lambda_{0}$ is the transition center wavelength, and $m$ is the mass in atomic units \cite{phillips2023}. The resulting kinetic temperatures for the neutrals and ions are plotted along with the excitation temperatures in Figure~\ref{Temperatures}. Except for at the earliest times, the kinetic temperatures for neutrals and ions are similar, both showing a decrease with time as expected from LPP cooling. For U~II, before 4.5~$\mu$s the excitation temperatures are notably higher than the kinetic temperatures, although the excitation temperatures drop rapidly. The U~II kinetic and excitation temperatures follow similar trends for most times, although at times from t $\sim$10-40~$\mu$s, the kinetic temperature is slightly lower. From t $\sim$15-50~$\mu$s, the U~I kinetic temperature appears significantly lower than the excitation temperature, and a similar behavior was observed previously for U~I under similar ablation conditions \cite{phillips2023}. The results suggest that the excitation and kinetic temperatures of neutral and ionized atoms may not be equal during these times of LPP evolution, which could indicate departures from LTE. However, spatial variations in temperature/density along the measurement path may also influence the apparent temperatures.

The Lorentzian component of the measured linewidth arises primarily from collisional broadening with electrons (Stark) and neutral atoms (van der Waals). It is expected that Stark broadening will be dominant at early times when electron densities are high, but van der Waals broadening will be dominant at later times; however, in the current experiments both broadening mechanisms should be considered for all times. Natural, resonance, and power broadening were neglected as they were estimated to be small relative to the other broadening mechanisms. Van der Waals broadening was calculated using the following equation:

\begin{align}
\centering
    \Delta\nu_{v} = \gamma_{ref} P \sqrt{\frac{T_{ref}}{T_{k}}}
\label{VanDerWaals}
\end{align}

\begin{figure}
\centering
\includegraphics[width=0.45\textwidth]{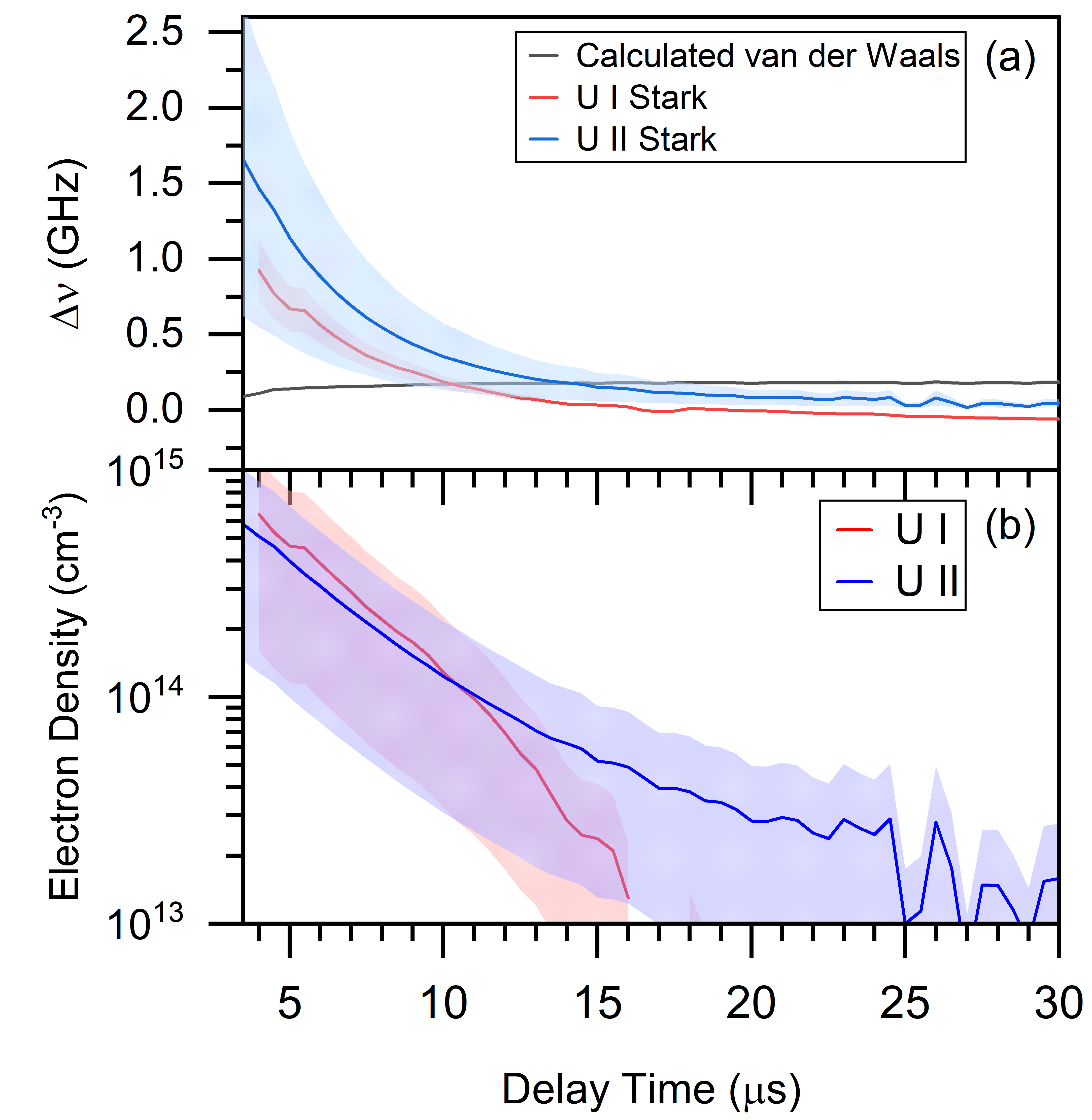}
\caption{\small{(a) Collisional broadening contributions to Lorentzian linewidths. The black trace shows the estimated van der Waals broadening (black) for U~I and U~II based on a broadening coefficient of 2.1~MHz/Torr (at 5300~K) and the measured average kinetic temperature. The blue (U~II) and red (U~I) traces show the estimated Stark linewidth obtained by subtracting the van der Waals linewidth from the measured average Lorentzian widths. (b) Resulting electron density as a function of time based on the Stark linewidths of U~I (red) and U~II (blue) transitions from (a). Shaded regions show the estimated uncertainties.}}
\label{Electron Density}
\end{figure}

Where $\gamma_{ref}$ is the pressure broadening coefficient, P is the experimental pressure, $T_{ref}$ is the reference temperature associated with the pressure broadening coefficient, and $T_{k}$ is the kinetic temperature of the measurement \cite{HariAPR2018}. Pressure broadening coefficients are not available in the literature for the U~I and U~II transitions measured here. We assumed $\gamma_{ref}$~=~2.1~MHz/Torr at $T_{ref}$~=~5300~K, as reported by Taylor and Phillips for a different U~I transition broadened by air \cite{taylor2014differential}. Figure~\ref{Electron Density}(a) shows the estimated van der Waals broadening, which ranges from $\sim$0.1-0.2~GHz due to the change in kinetic temperature versus time. The estimated van der Waals broadening is also seen to contribute a significant fraction of the total Lorentzian linewidths shown in Fig.~\ref{Voigt Gaussian Lorentzian}, and is approximately equal to the minimum average Lorentzian FWHM at late times within uncertainties. 

To estimate the Stark broadening contribution to the linewidth, the calculated van der Waals linewidth was subtracted from the average Lorentzian linewidth at each time step, and the results are shown in Fig.~\ref{Electron Density}(a). Uncertainty bands were calculated based on the uncertainties in Lorentzian and van der Waals linewidths. Fig.~\ref{Electron Density}(a) shows that the estimated Stark broadening decays toward zero at later times, indicating that the pressure broadening coefficient used to estimated the van der Waals broadening was reasonable, at least relative to other uncertainties. 

Electron density was estimated using the standard relationship between Stark broadening and electron density \cite{2022-Hari-RMP}:

\begin{align}
\centering
    \Delta\lambda_{S} = 2W\left(\frac{N_{e}}{10^{16}}\right)
\label{Stark}
\end{align}

where $W$ is the Stark broadening (electron-impact) parameter, and $N_{e}$ is the electron density. 
Temperature-dependence and ion-broadening contributions have been neglected as they are expected to be small relative to the uncertainties in the current calculations. As with van der Waals broadening, Stark broadening parameters were not available in the literature for the measured U~I and U~II transitions. Therefore, we assumed average Stark broadening parameters of 6~pm (U~I) and 12~pm (U~II), as measured for alternate uranium transitions by Burger et al. \cite{BurgerPoP2019}. To account for potential variations in Stark parameters between transitions, we estimated they could range from 0.5-2 times the average value used. Figure~\ref{Electron Density}(b) shows the resulting electron density calculated from the estimated Stark broadening. The electron densities based on the measured U~I and U~II transitions are in good agreement from 3.5-15~$\mu$s, at which time the U~I Stark linewidth becomes indistinguishable from zero within uncertainties. The electron density from U~II transitions could be determined until $\sim$25~$\mu$s. Overall, the electron density ranged from $\sim$5$\times$10$^{14}$ - 2$\times$10$^{13}$~cm$^{-3}$ over times 3.5-25~$\mu$s. Despite the large uncertainties arising from both the measurements and in the Stark/van der Waals broadening parameters, the results give an order-of-magnitude estimate of electron density which also shows the expected decrease in magnitude over these time scales.

\section{Discussion}

The electron density estimated from Stark broadening of absorption spectra provides important information on later times of LPP evolution not typically available via OES/LIBS methods. Many OES measurements of electron density in the early stages of LPP evolution are reported in the literature, with a few being performed on uranium LPPs. In Singh et al., the electron density of a uranium LPP in 1~atm Ar measured using OES with Saha-Boltzmann analysis was 
$\sim$1$\times$10$^{16}$~cm$^{-3}$ at 10~$\mu$s, decaying to $\sim$5$\times$10$^{15}$~cm$^{-3}$ at 20~$\mu$s \cite{SinghPST2018}. Skrodzki et al. measured electron densities of $\sim$5$\times$10$^{16}$~cm$^{-3}$ at 5~$\mu$s for ablation of a uranium-doped glass in 100~Torr nitrogen \cite{SkrodzkiSCAB2016}. Burger et al. found electron densities using OES/Stark broadening in a uranium LPP in 5~mTorr nitrogen to be $\sim$1$\times$10$^{16}$~cm$^{-3}$ at 1~$\mu$s \cite{BurgerPoP2019}. Hull et al. measured an electron density of $\sim$3$\times$10$^{16}$~cm$^{-3}$ at 10~$\mu$s using OES of a uranium LPP in atmospheric pressure air \cite{hull2022combined}. The electron densities estimated here using LAS from 3.5-25~$\mu$s in 50~Torr Ar are $\sim$~1-2 orders of magnitude smaller than the values reported using OES/LIBS, which is reasonable considering the OES measurements are mostly for earlier times. However, it appears that the LAS-based electron density is lower than comparable OES electron densities obtained at similar times, which has been noted previously \cite{Gornushkin1999}. It is possible that the absorption-based method is more sensitive to cooler regions of the LPP with a corresponding lower electron density, while OES methods are more sensitive to higher temperature regions, based on spatial averaging effects \cite{Aguilera2004}. However, considering differences in ablation conditions between the various measurements and other experimental uncertainties, it is difficult to draw conclusions at this time and further study is needed. Previous measurements using frequency comb absorption spectroscopy found electron densities of $\sim$10$^{14}$~cm$^{-3}$ from 20-50~$\mu$s based on Saha-Boltzmann analysis of a Fe LPP in 100~Torr Ar \cite{mccauley2024dual}, and the results here show a similar order-of-magnitude.

The U~II column density could be measured to times beyond which the Stark linewidth could be estimated, and this can be used to estimate a lower limit of electron density assuming charge neutrality and spatial uniformity of the LPP \cite{mccauley2024dual}. Assuming a LPP diameter of $\sim$1~cm, the U~II column density shown in Fig.~\ref{Density and Temp}(a) indicates a minimum electron density of $\sim$2$\times$10$^{15}$~cm$^{-3}$ at 5~$\mu$s, $\sim$1$\times$10$^{15}$~cm$^{-3}$ at 10~$\mu$s, and $\sim$2$\times$10$^{14}$~cm$^{-3}$ at 70~$\mu$s. These values are higher than the electron densities determined from the LAS Stark broadening, which may indicate a larger effective size for the LPP or spatial non-uniformity in U~II and/or electron density along the measurement path. It is also possible that the Stark broadening coefficient used in the LAS calculation was overestimated, or that the Stark linewidth was underestimated. Inclusion of other potential line-broadening mechanisms (e.g. resonance broadening, power broadening), would increase the discrepancy between U~II and electron densities. Further work is necessary to identify and understand the source of these apparent differences in electron density, and particularly to measure whether a spatial inhomogeneity between ion and electron densities may exist even at these relatively late times in LPP evolution. 

The results provide insight into the ionization states of LPPs at later times in their evolution. We measured a ratio of U ion to neutral column density of 14.42 at 15~$\mu$s. Weerakkody et al. measured a U II to U I ratio of $\sim$10 at 10~$\mu$s when ablating a U$_{2}$Si$_{3}$ target into a 15~Torr mixture of 2\% O$_{2}$ and 98\% Ar \cite{WeerakkodySCAB2020}. Merten et al. measured ion and neutral total number densities for Y in various background gases at a pressure of 225~Torr (300~mbar) using absorption spectroscopy, and also found that the LPP was highly ionized at early times; for example, the ion to neutral ratio was $>$1 for t $\lesssim$ 6~$\mu$s in Ar \cite{merten2023time}. McCauley et al. measured an ion to neutral column density ratio to be $\sim$1 at 10~$\mu$s for a Fe LPP in 100~Torr Ar \cite{mccauley2024dual}. Accounting for differences in ablation conditions and ionization potentials of different elements, the results obtained here appear consistent with these prior results. The results show that under these ablation conditions uranium may remain highly ionized in the LPP for tens of microseconds, and U~II absorption from some transitions could be detected until 110~$\mu$s. These results are important in regard to line selection for uranium isotopic measurements, since U~II transitions have larger isotope shift on average than U~I transitions \cite{HariAPR2018}. The ionization state of uranium is also important for predicting chemical reaction kinetics, which depend on the charge states of the reactants \cite{Kautz_2021_UReview}.

Using the estimated electron density (1$\times$10$^{14}~$cm$^{-3}$) and average U excitation temperature of 3,000~K at 15~$\mu$s, the Saha equation \cite{kunze2009} predicts the expected ratio of U~II to U~I number density to be $\sim$5$\times10^{-4}$, which is significantly different than the ratio of $\sim$14 obtained from measured column densities. The Saha equation is highly sensitive to both electron density and temperature. For example, a U~II/U~I ratio of 14 could be obtained for an electron density of $\sim$1$\times$10$^{14}$~cm$^{-3}$ and a higher electron temperature of 5,000~K, or for a lower electron density of $\sim$1$\times$10$^{10}$~cm$^{-3}$ at a temperature of 3,000~K. Thus, if the Saha equation applies under the measured LPP conditions, it appears the electron density may be overestimated or the temperature is underestimated. It is important to note that the U~II/U~I column density ratio may not be equal to the number density ratio if a spatial inhomogeneity in ion, neutral, and electron densities is present and this may explain some of the discrepancy. Nevertheless, the measurement suggests that the ionization in the uranium LPP is much higher than the Saha equation would predict, which has also been noted in other absorption measurements in U LPPs \cite{WeerakkodySCAB2020}. 

The apparent inconsistencies between average electron density, ionization ratio, and temperatures highlight some of the challenges with measurement of non-stationary plasma sources such as LPPs. It has previously been noted that LTE assumptions may not be valid at all times in LPP evolution \cite{kunze2009, Hahn2010}; thus, the electron and atomic excitation temperatures may be different and the Saha ionization equation may not apply. According to the often applied McWhirter criterion \cite{kunze2009, Hahn2010}, the electron density should be $\gtrsim$10$^{14}$~cm$^{-3}$ under these LPP conditions to ensure LTE is established, and the estimated electron density is not significantly above this threshold. The observed differences between excitation and kinetic temperatures in Fig.~\ref{Temperatures} indicate that the LPP system is not described by a single temperature \cite{Gornushkin1999}. It is interesting to note that the U~I and U~II Boltzmann plots both showed good linearity at all times, and the kinetic temperatures for different measured lines were in good agreement with each other, all of which suggest that a thermal equilibrium may be established for energy levels within a given species, even if thermal equilibrium is not present between different species. However, the results also suggest that a Saha-Boltzmann analysis of the U~I and U~II transitions may not be appropriate for these times in the LPP.

Aside from uncertainties in spectroscopic parameters (oscillator strengths, Stark and van der Waals broadening coefficients, etc.), the largest source of uncertainty in the measured parameters (temperatures, U~I/U~II column densities, and electron density) likely arises from potential spatial inhomogeneity of the LPP along the measurement path. It has been observed in OES/LIBS experiments that not accounting for spatial gradients using an Abel inversion can lead to differences in apparent ion/neutral temperatures \cite{Aguilera2004}. For the times measured here with LAS, the spatial gradients are expected to be smaller than in the early stages of LPP expansion measured with OES, and prior results have shown that the LPP expansion has stopped by $\sim$5-10~$\mu$s under similar ablation conditions \cite{HariJAP2022}. Nevertheless, potential variations in temperatures and species along the measurement path should be considered when interpreting and comparing results. For example, nonuniform temperatures along the path may lead to deviations from Voigt lineshapes. Similarly, periods of rapid LPP expansion may lead to non-thermal velocity distributions of atoms that also cause deviations from Voigt lineshapes. For ablation in low pressure conditions, clear Doppler splitting has been observed in LAS measurements \cite{bushaw1998investigation, Miyabe2012Ce}. Although Doppler splitting was not observed in the current experiment, it is possible that unresolved non-thermal Doppler effects may be present for spectra measured at the early times.

\section{Conclusions}
\label{conclusions}

In summary, high-resolution and time-resolved measurements of ten neutral and six singly-ionized transitions of uranium atoms were used to measure properties of ionization, electron density, and temperature in a LPP created under 50~Torr Argon. Uranium ionic absorption was detected in the LPP for $>$100~$\mu$s after plasma formation, while absorption from neutral transitions was observed for $>$1,000~$\mu$s. The ionization fraction U~II/U~I was greater than 10 at 15~$\mu$s and greater than unity until 50~$\mu$s, which deviates from predictions based on the Saha equation. Spectral fitting extracted Stark broadening of ion and neutral absorption lines, which was used to estimate electron densities ranging from $\sim$10$^{15}$-10$^{13}$~cm$^{-3}$ over times 4-25~$\mu$s. Differences between excitation and kinetic temperatures were also observed, indicating possible departures from LTE conditions as the LPP cools. The results provide valuable new information on charge properties, electron density, and temperatures in uranium LPPs at later times which are difficult to measure using optical emission methods.

It is worth noting that the U~II transitions in the near-infrared spectral region measured here have relatively low oscillator strengths compared to transitions in the visible or near-ultraviolet regions. Future measurements of these stronger ionic transitions could allow better determination of U~II excitation temperatures and total column densities over a wider range of energy levels, and extend measurements to later times. It may also be possible to measure higher ionization states of U in this spectral region. However, the high density of transitions in the visible/ultraviolet increases the likelihood of overlapping blended lines, which makes spectral fitting more challenging \cite{phillips2023}. 

The same experimental methods used here could be extended to elements other than uranium. While the availability of oscillator strength data for uranium is advantageous for the current studies, the lack of information on Stark and van der Waals broadening led to high uncertainties in electron density. It would be beneficial to perform similar LPP measurements using LAS of elements having transitions with known Stark broadening coefficients, and comparing results with OES measurements, to evaluate uncertainties in determination of electron density. Results could also be compared with Stark shift measurements, and consistency between Stark shift/broadening can be useful for discriminating it from other broadening processes \cite{Gornushkin1999}. For the measurements here, shifts in the center frequencies of the transitions were observed, but could not be attributed to Stark shifts.

\section*{Acknowledgments}
This work was supported by the Department of the Defense, Defense Threat Reduction Agency (DTRA) under award number HDTRA1-20-2-0001. The content of the information does not necessarily reflect the position or the policy of the federal government, and no official endorsement should be inferred. Pacific Northwest National Laboratory is a multi-program national laboratory operated by Battelle for the U.S. Department of Energy under Contract DE-AC05-76RL01830.

\section*{Declaration of Competing Interests}
The authors declare that they have no known competing financial interests or personal relationships that could have appeared to influence the work reported in this paper.

\bibliography{text.bib}
\bibliographystyle{elsarticle-num-names.bst}

\end{document}